\documentclass[preprint2]{aastex62}

\usepackage{amsmath}
\usepackage{natbib}
\usepackage{graphicx}
\usepackage{hyperref}

\newcommand{\astropop}{\textsc{astropop}}
\newcommand{\pccdpack}{\textsc{pccdpack}}

\newcommand{\iraf}{\textsc{iraf}}

\received{June 5, 2018}
\revised{September 5, 2018}
\accepted{October 27, 2018}

\submitjournal{PASP}

\shorttitle{\astropop}
\shortauthors{Campagnolo, J. C. N.}

\begin{document}

\title{\astropop: the ASTROnomical POlarimetry and Photometry pipeline}

\correspondingauthor{Julio Cesar Neves Campagnolo}
\email{juliocampagnolo@on.br}

\author[0000-0002-2075-2424]{Julio Cesar Neves Campagnolo}
\affil{Observat\'{o}rio Nacional, 20921-400, Rio de Janeiro, Brazil}

\begin{abstract}

We developed a new pure-python pipeline to reduce photometric and polarimetric data: \astropop. It has been designed and optimized to work fully automated with the IAGPOL polarimeter of Pico dos Dias observatory (OPD, Brazil) and can reduce photometry and polarimetry data from other instruments, especially from SPARC4, a multi-channel polarimeter that has been developed for OPD. We present the results produced by this new code, and compare them with those obtained from \pccdpack, a traditionally used \iraf\ package developed for IAGPOL. We also propose to use this code for automatic photometric reduction for the new ROBO40 telescope, also installed at OPD. \astropop\ is fully open source and distributed under the BSD-3 clause license.

\end{abstract}


\section{Introduction} \label{sec:intro}

In the last decade, the optimization of astronomical instrumentation and the advent of fast CCD cameras have generated a large increase in the amount of data that can be produced by just one telescope, even a small one. The reduction of all this data can be a bottleneck that slows down the science production and can consume a large fraction of an astronomer's productive time.

Automatic pipelines to reduce data are common for big observatories, but small observatories generally lack this type of support or, if it exists, the reduction codes strongly need human interaction. For example, one of the most used instruments at Observat\'orio Pico dos Dias (OPD, Brazil) is the IAGPOL polarimeter \citep{1996ASPC...97..118M}, which is portable and can be used in the three main telescopes of the observatory. This instrument already has a reduction pipeline, called \pccdpack\ \citep{2000PhDT........82P}. This reduction software is well-tested and highly reliable, being the standard reduction code for this instrument. However, this software needs significant user interaction and becomes impractical for surveys and other large datasets.

This will be even more critical for the new polarimeter, Simultaneous PolArimeter and Rapid Camera in 4 bands \citep[SPARC4, ][]{2012SPIE.8446E..26R}, which is being developed and will replace the IAGPOL at OPD's 1.6\,m telescope. SPARC4 will be equipped with 4 rapid EMCCD cameras (instead of just one like in IAGPOL), producing a greater amount of data.

Recently, a new pipeline called \textsc{solvepol}, was developed by \citet{2017PASP..129e5001R} to replace \pccdpack. This new package is written in \textsc{interactive data language} (\textsc{idl}), and it was created to work in an automated way for the SOUTH POL survey \citep{2012AIPC.1429..244M}. However, \textsc{idl} is a paid license software, making portability to and usability in other projects a possible problem.

Another new critical instrument installed in OPD is the ROBO40 telescope. A small robotic telescope with 40\,cm of diameter with photometric proposes, still in commissioning phase. This instrument will perform automatic observations and will produce a large amount of data every night, but at this time, lacks an automatic reduction script ready at now.

I have developed a new code \astropop: the ASTROnomical POlarimetry and Photometry pipeline 
\citep{2018ascl.soft05024C}, which is modular, automatic and easily used to reduce IAGPOL and ROBO40 data. This code is designed to be used for photometric or polarimetric data reduction, without user interaction or user programming. It is written in Python, an open source language largely used by the astronomical community for data reduction and analysis. Having only dependencies of python packages, \astropop\ can be installed under Python environment with version newer than $3.5$. It can be installed automatically by PyPi\footnote{\url{https://pypi.org/}} or Anaconda\footnote{\url{https://www.anaconda.com/}}. The code can be accessed at \url{https://github.com/juliotux/astropop} and its documentation is available in the ReadTheDocs\footnote{\url{https://astropop.readthedocs.io}} platform. For distribution, I choose the BSD-3 clause license\footnote{\url{https://opensource.org/licenses/BSD-3-Clause}}, which is very permissive and allow any modification, replication and use (including commercial).

In this work we will describe the reduction algorithms of \astropop\ and compare its results with those ones from \pccdpack\ and form the literature. In Sect.~\ref{sec:description}, we discuss the code and the process made in each reduction step. In Sect.~\ref{sec:observation}, we describe the observations used for tests, which are showed and discussed in Sect.~\ref{sec:results}.

\section{Code algorithms and description\label{sec:description}}

\begin{figure}
\centering
\includegraphics[width=\linewidth]{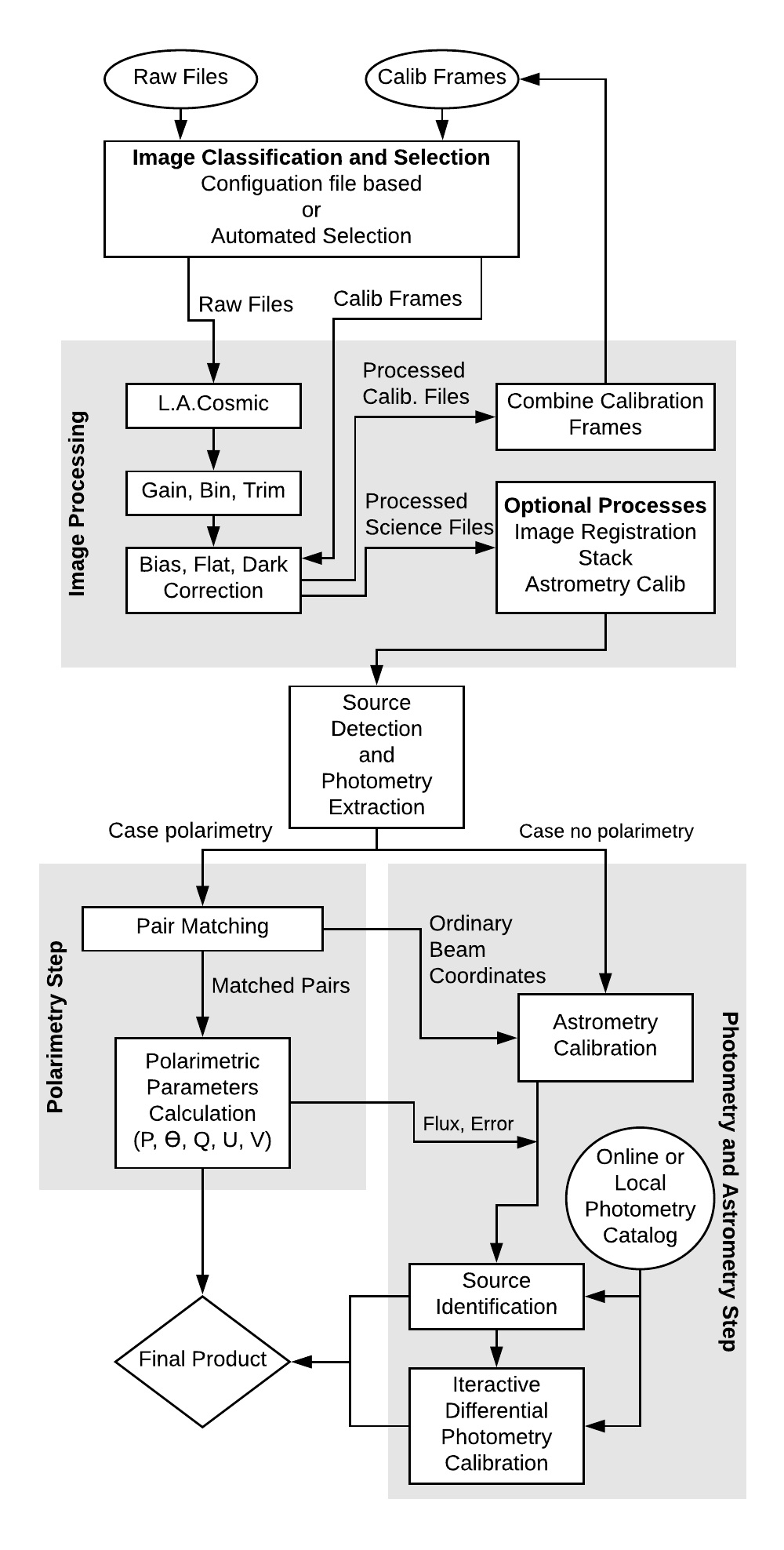}
\caption{Schematic recipe for IAGPOL and ROBO40 reductions, implemented in \astropop.\label{fig:steps}}
\end{figure}

\astropop\ is a data reduction pipeline written in Python~3 language and designed to perform the standard reduction process for polarimetry or photometry. The pipeline is divided in different modules, each one with a specific function, but working in an homogeneous and integrated way. Fig.~\ref{fig:steps} summarizes every step of the IAGPOL and ROBO40 reduction recipes using the \astropop\ modules, described in this section.

As internal image storage format, the code uses the standard Astropy\footnote{\url{https://astropy.org}} FITS Image HDU class, which makes any module of the code easy to integrate with another Python code based in Astropy.

\subsection{Image Pre-processing}

The image pre-processing step performs the basic image reduction, including bias or dark subtraction, flat-field correction, image trimming and binning, cosmic ray extraction, and gain correction. Image alignment, trimming to the target shape and image coaddition with sigma and extrema data clippings and image normalization tasks are also available. The code can also create master calibration frames (like master bias, master flat, and bad pixel mask), based on raw images when needed.

The first reduction step is the cosmic ray extraction, with the \textsc{astroscrappy}\footnote{\url{https://github.com/astropy/astroscrappy}} package, based on Laplacian edge detection through the L.A.Cosmic algorithm \citep{2001PASP..113.1420V}.

After that, bias, flat, dark, trimming, binning and coaddition task are performed using built-in functions, based on our Python implementation of the \iraf\ tasks \textit{imarith} and \textit{imcombine}. These tasks are all optional and will only be executed if the user set the correct parameters. The program can also check header keywords to determine if the correct images have been used in the process.

Another important feature of the code is the astrometric alignment and combination of datasets. When the data to be reduced is composed by several images that have to be combined, it is common that small shifts, produced by the bad tracking of the telescope during the observation, interfere in the results. To fix this problem, \astropop\ can align the images by two different ways: (i) by registering the Fourier transform correlation of the images, or (ii) by asterism matching of detected sources.

The Fourier transform correlation is performed by the \textsc{scikit-image} \footnote{\url{http://scikit-image.org/}} function \textsc{register\_translation}, which uses the optimized algorithm described by \citet{2008OptL...33..156G}, based on the cross-correlation of the images in the Fourier space. This code is only applicable for images with the same scale and rotation, because it has the limitation of only finding shifts due to image translation. The alignment precision is of the order of a half pixel, since the shifts are calculated in integer pixel values.

For images where sources can be detected, the user can also choose the alignment of the images by asterism matching, using the \textsc{astroalign}\footnote{\url{https://github.com/toros-astro/astroalign}} package. It forms groups of three stars (asterisms) from a source-list and finds the best matching of the asterism in a target list, finding the necessary transformation to align the images. This process allows astrometric alignment with subpixel precision, limited by the centroid detection precision and the image quality.

\subsection{Source detection and photometry}

There are several open source Python packages that perform reliable CCD photometry. \astropop uses the \textit{Source Extraction and Photometry} \citep[\textsc{sep}, ][]{2016JOSS....1...58B}, which is a implementation of the \textsc{sextractor} algorithm \citep{1996A&AS..117..393B} in Python and \textsc{daophot} \citep{1987PASP...99..191S} based code from the \textsc{idl-astrolib}\footnote{Translation originally done by \citet{2015ascl.soft01010J} for the \textsc{pythonphot} code and updated for \astropop.} \citep{1993ASPC...52..515V}.

\astropop\ source detection consists of two main algorithms: image segmentation detection directly from \textsc{sep}; and the source finding algorithm from \textsc{daophot} ported to \astropop. For an optimal automated extraction of point sources, I also developed a function that uses both codes: first it detects all the possible sources using \textsc{sep}, then calculates the FWHM of the PSF from Moffat or Gaussian fitting and use this parameter to better extract sources using the \textsc{daophot} algorithm.

The aperture photometry is also from \textsc{sep}, except that the local background subtraction is implemented directly in \astropop. \textsc{sep} performs this subtraction based on the average of the pixels inside the background annulus, without any cleaning or source masking. \astropop\ performs a sigma clipping in the background annulus, in order to mask the contribution of other sources, and obtains the sky value by calculating the median of the annulus pixels or by the MMM mode estimator from \textsc{daophot} ($3\times\mathrm{median}-2\times\mathrm{mean}$).

PSF photometry is planned for the next version of the code.

\subsection{Polarimetry reduction}

The \astropop\ code is primarily developed to deal with dual-beam polarimeters, which are useful for point sources. This kind of instrument, as described in \citet{1996ASPC...97..118M}, consists of a retarder plate which modulates the incoming polarization, an analyzer that splits the incoming light into two beams with orthogonal polarization; and the detector. Currently, the code requires additional modification to handle single-beam polarimetry, where the analyzer is a polaroid sheet that filter the light and allows just one polarized beam to reach the detector. This mode is used for extended sources and is also available on IAGPOL. A proper handling for this polarimeters is planned to future versions. Here we will focus just on the reduction of the dual-beam mode.

\begin{figure}
\includegraphics[width=\linewidth]{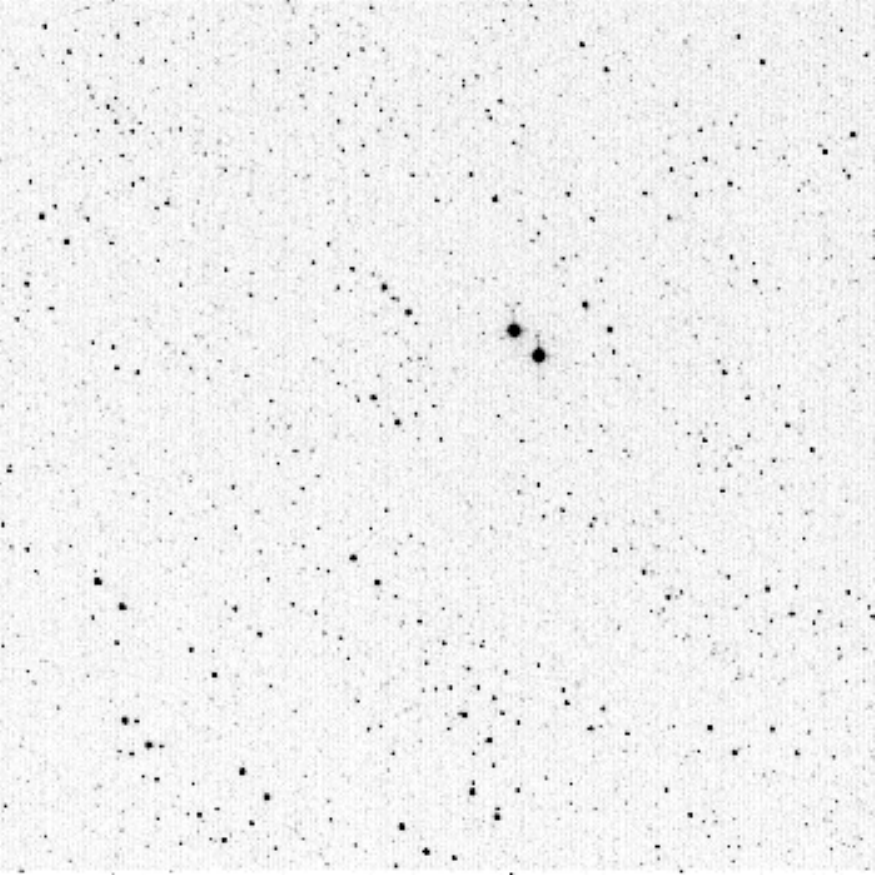}
\caption{Example image obtained from a dual beam polarimeter. The figure shows the field of the \object{HD\,172252} star from IAGPOL, where all stars appear doubled, as result of from the superposition of the images formed by the ordinary and extraordinary polarized beams.\label{fig:dual_beam}}
\end{figure}

As mentioned, a dual-beam polarimeter splits the light from the telescope in two beams, which form two superposed images shifted by some fixed distance in the detector. Fig.~\ref{fig:dual_beam} shows an example image obtained from the IAGPOL polarimeter with the two orthogonal polarized images superposed. The first step of the polarimetry reduction is to identify this shift and match the pair of sources that corresponds to each star.

To compute the shift between the images formed by the ordinary and extraordinary beams, we calculate the distance in $x$ and $y$ image coordinates between all the sources in the image -- i.e. the distance between the stars in all possible pair-combinations in the image -- and them we find the most common distance in each coordinate using the peak of the histogram of the distances. To improve the precision of the distance determination, the code iteratively clips the histogram around the found peak, improving the definition of the bins in that region and finding the a more precise peak value. When the shift is identified, the sources are grouped in the ordinary/extraordinary pairs. As convention, the source with lower $y$ coordinate in the pair is considered the ordinary one.

Polarimetry reduction is possible only for sets of images in the same field, with different rotation positions of the retarder plate. In the case of linear polarimetry, a half-wave retarder plate is used and it is necessary at least 4 images rotated by 22.5$^\circ$ degrees each one to have a reliable value.

The reduction process consists of the measurement of the flux of the ordinary and extraordinary beams in all images of each star. We then proceed to compute the relative difference between these fluxes, using equation using:

\begin{equation}\label{eq:z}
z(i) = \frac{F_i^o - F_i^e\cdot k}{F_i^o + F_i^e\cdot{k}},
\end{equation}

\noindent where $z(i)$ is the difference between the ordinary ($F_i^o$) and extraordinary ($F_i^e$) fluxes in an image $i$, and $k=\frac{\sum_i F_i^o}{\sum_i F_i^e}$ is a normalization constant to correct for a possible different response of the instrument to the ordinary and extraordinary polarizations. The $z(i)$ factor is related to the Stokes $Q$ and $U$ parameters of polarization, according to equation\,\ref{eq:stokes}, which is fitted to the data by the code using the least squares minimization and the Levenberg--Marquardt optimization, in order to obtain $Q$ and $U$ values. This method will be called here as Stokes Least Squares (SLS).

\begin{equation}\label{eq:stokes}
z(i) = Q\cos(4\Psi(i)) + U\sin(4\Psi(i)),
\end{equation}

\noindent where $\Psi(i)$ is the rotation of the retarder plate in each position. In addition, $Q$ and $U$ are related to the physical polarization level $P$ and the position angle of the polarization $\Theta$ by the relations in the equations \ref{eq:p} and \ref{eq:t}:

\begin{eqnarray}
\label{eq:p}
P &=& \sqrt{Q^2 + U^2}, \\
\label{eq:t}
\Theta &=& \frac{1}{2}\arctan{\left(\frac{U}{Q}\right)}.
\end{eqnarray}

The errors estimated for the $Q$ and $U$ are obtained from the diagonal of the covariance matrix of the fitting, and are related to the errors of $P$ and $\Theta$ according to:

\begin{eqnarray}
\label{eq:errp}
\sigma_P &=& \frac{1}{P}\sqrt{(\sigma_Q Q)^2 + (\sigma_U U)^2}, \\
\label{eq:errt}
\sigma_\Theta &=& 28.65^\circ \cdot \frac{\sigma_P}{P}.
\end{eqnarray}

Alternatively, to compute $Q$ and $U$, \astropop\ also has the algorithm described by \citet{1984PASP...96..383M}, which will be called here `MBR84' and is the same used by \pccdpack\ and \textsc{solvepol}. With this algorithm, $Q$ and $U$ are obtained from the following equations:

\begin{eqnarray}
\label{eq:qsum}
Q &=& \frac{2}{n}\sum_{i=1}^{n} z(i) \cos{4\Psi(i)}, \\
\label{eq:usum}
U &=& \frac{2}{n}\sum_{i=1}^{n} z(i) \sin{4\Psi(i)},
\end{eqnarray}

\noindent where $n$ is the total number of retarder positions observed, $\sigma_P$ is computed as:

\begin{equation}
\sigma_P = \sqrt{\frac{1}{n-2} \left(\frac{2}{n} \sum_{i=1}^{n} z(i)^{2} - P^2\right)}.
\end{equation}

The polarization level is intrinsically positive (there is no negative polarization) and has a Rice error distribution \citep{2006PASP..118.1340V}. This produces an effect of Ricean bias in the polarimetric measurements that affects mainly the values with $P/\sigma_P<3$ \citep{1986VA.....29...27C}. \astropop\ do not perform any kind of correction of the Ricean bias. If the users want to use data with low signal quality, they have to perform the correction by themselves.

\subsection{Astrometric and photometric calibration}

Astrometric calibration is performed with the source coordinates taken from the source-detection process. It can be calculated by two different methods: (i) by resolving the field with the \textsc{astrometry.net} software \citep{2010AJ....139.1782L}, or (ii) by manual setting of the brightest star coordinates and field orientation. The coordinates used are those from the ordinary beams in the image.

For the \textsc{astrometry.net} calibration, the code creates a list of stars sorted by brightness, which is provided to the program that will search in specific index files for the best solution of matching asterisms formed by 4 stars. This process can be time consuming, but the \astropop\ can optimize the processing time to just a few seconds by looking in the image header for the approximate field center coordinates and plate scale.

The second astrometric calibration method can be used to calibrate the images when the \textsc{astrometry.net} calibration fails or is impossible due to the low number of detected stars. In this method, the user provides the coordinates of the brightest star, the plate scale and the field orientation to the code, which makes the astrometry calibration based on the coordinates of this star.

Through the astrometry calibration, the stars can be identified in on-line catalogs. With this information, the code collects their available magnitudes, enabling the calibration of the extracted photometry. To do this, the code converts the measured flux into instrumental magnitudes and computes a correction factor (zero point), that will be used to convert the instrumental magnitudes to the calibrated standard magnitudes.

The zero point is computed based on the difference of the catalog and the instrumental magnitudes for all the stars in the field. The final value for the field can be obtained by the median of the individual stars zero point, or by an iterative median from a Monte-Carlo algorithm.

This Monte-Carlo algorithm makes several iterations where a random group of stars is chosen in field and a partial zero point is computed for that iteration from the median of the individual chosen stars. The zero point is computed as the median of all the partial values from each iteration.

This final correction factor is them added to the instrumental magnitudes for each star in order to obtain the final calibrated magnitude value. The final errors are estimated as the sum of the standard Poisson photon noise and the standard deviation ($1\sigma$) of the zero points calculated for each star (in the case of median zero point) or for each iteration (in the case of Monte-Carlo algorithm). Calibration errors, as determined by differences bright star magnitudes, typically range between 0.03 and 0.1 magnitudes, although errors as low as 0.01 magnitudes for fields with many well-exposed stars.

Currently, there are six standard on-line photometric catalogs that can be used for the star identification and zero point calibration. One of these catalogs is the direct access to the stellar informations from the Simbad database \citep{2000A&AS..143....9W}. The other five are accessed via Vizier platform: UCAC4 \citep{2012yCat.1322....0Z, 2013AJ....145...44Z}, 2MASS \citep{2003yCat.2246....0C}, DENIS \citep{2000A&AS..144..235C, 2005yCat.2263....0T}, APASS~DR9 \citep{2015AAS...22533616H, 2016yCat.2336....0H} and GSC2.3 \citep{2008AJ....136..735L}. In addition, custom ASCII local catalogs can be used. The filter matching is configured in the catalog declaration and is based on column names. None of these catalogs have color corrections enabled at the moment, however, the \astropop\ API allow it to be set for custom catalogs. For the ROBO40 and IAGPOL recipes, we use the APASS~DR9 catalog for the $B$ and $V$ filters calibration, the GSC2.3 for the $R$ filter and DENIS catalog for the $I$, which are the main filters available for these instruments.

\section{Observations\label{sec:observation}}

In this work I used different datasets of standard polarized stars, obtained during six observing runs using the 0.6\,m Boller \& Chivens telescope at OPD between 2016 and 2017. In addition, older observations with the same instrument and telescope available at OPD databank\footnote{\url{http://lnapadrao.lna.br/OPD/databank/databank}} were used to expand the sample.
The telescope was mounted with the IAGPOL polarimeter \citep{1996ASPC...97..118M}, which consists of a retarder plate, followed by a Sarvat prism, a filter wheel and a Andor IkonL CCD detector with 2048$\times$2048 of 13.5\,\micron\ square pixels. The detector was configured in 2$\times$2 pixels binning, giving a plate scale of 0.64\arcsec/bin, or $\sim11\arcmin\times11\arcmin$ of field of view. The observations from the databank taken in 1998 used the OPD's CCD\,101, with $1050\times1050$ square pixel and a plate scale of 0.57\arcsec/pixel. For these observations, no binning was used.

The instrumental setup of the polarimeter splits the incoming light in two beams with orthogonal polarization (ordinary and extraordinary) and forms two images in the detector that are shifted from each other by some pixels. To cover a large range of magnitudes, two different Sarvat analyzers were used: the ``A2'' calcite, with a neutral density filter of 1.2\,mag, that produces beam shifts of $\Delta(x,y)=(4, 33)$\,bins, that corresponds to a separation of $21\arcsec$; and the ``A0'' calcite, with no density filter, which produces image shifts of $\Delta(x,y)=(28, 28)$\,bins or $25\arcsec$.

In our observations, we used the IAGPOL equipped with the half-wave retarder plate and, for each dataset, 16 acquisitions of the field were done, each one with the retarder plate rotated 22.5\arcdeg\ in relation to the previous one, totaling a full $360^\circ$ rotation. 

For the reduction with \astropop, we used a fixed aperture of 4 bins. The sky subtraction annulus was defined with an inner radius of 10 bins and a width of 10 bins, to ensure the source pairs are not included. The same aperture and sky annulus were used with \pccdpack\ for the comparison of our results. Both codes are capable of automatically selecting the aperture where the best polarimetric SNR was found for each star for the main output catalogs, however, I opted for a fixed value for a more direct comparison.

\section{Results and Discussions\label{sec:results}}

To check the reliability of the results produced by \astropop, I compared the obtained and simulated data with the literature and with the \pccdpack\ results. 

\subsection{Standard stars from literature}

We used the code to reproduce results from polarimetric standard stars from the literature, which are listed in Table\,\ref{tab:obs_stars}.

\begin{deluxetable*}{cCCcCDD}
\tablecaption{Catalog of literature values and observations of standard-polarized stars used in this work. For the observations, the results presented were reduced by \astropop\ using SLS algorithm. The $I$ polarization reference value for HD\,111579 star was extrapolated from $V$ value using \citet{1975ApJ...196..261S} relation. \label{tab:obs_stars}}

\tablehead{\colhead{Name} & \colhead{RA} & \colhead{Dec} & \colhead{Ref/Obs. Date} &
           \colhead{Filter} & \twocolhead{$P_V$} & \twocolhead{$\Theta_V$} \\
           \nodata & \colhead{HH:MM:SS} & \colhead{DD:MM:SS} & \colhead{YYYY-MM-DD} &
           \nodata & \twocolhead{\%} & \twocolhead{\arcdeg}}
\decimals
\tiny
\startdata
  \object{HD 14069}  & 02:16:45.2 & 07:41:10.7  &    $a$     & V & 0.022\pm0.019 &  156\pm24    \\
                     &            &             & 1998-11-25 & V &  0.16\pm0.08  &   \nodata    \\ \hline
  \object{HD 23512}  & 03:46:34.2 & 23:37:26.5  &    $b$     & V &  2.26\pm0.01  &  29.9\pm0.1  \\
                     &            &             & 1998-11-24 & V &  2.13\pm0.09  &   \nodata    \\
                     &            &             & 1998-11-25 & V &   2.1\pm0.1   &   \nodata    \\ \hline
  \object{HD 110984} & 12:46:44.8 & -61:11:11.6 &    $b$     & V &  5.70\pm0.02  &  91.6\pm0.1  \\
                     &            &             & 2015-06-21 & V &   5.9\pm0.3   &   \nodata    \\ \hline
  \object{HD 111579} & 12:51:03.6 & -61:14:37.7 &    $c$     & V &     6.290     &     101.7    \\
                     &            &             & $V$ extra. & I &     5.44      &   \nodata    \\
                     &            &             & 2015-05-21 & V &  6.02\pm0.2   &   \nodata    \\
                     &            &             & 2010-06-01 & I &  5.34\pm0.04  &   \nodata    \\ \hline
  \object{HD 126593} & 14:28:50.9 & -60:32:25.1 &    $b$     & V &  5.02\pm0.01  & 75.2\pm0.05  \\
                     &            &             & 2015-05-21 & V &  4.86\pm0.09  &   \nodata    \\ \hline
  \object{HD 145502} & 16:11:59.7 & -19:27:38.6 &    $c$     & V &     1.21      &    140.4     \\
                     &            &             & 2017-09-14 & V &  1.18\pm0.01  &   \nodata    \\ \hline
  \object{HD 147084} & 16:20:38.2 & -24:10:09.5 &    $b$     & V &  4.18\pm0.02  & 32.0\pm0.1   \\
                     &            &             & 2017-09-14 & V &  4.09\pm0.03  &   \nodata    \\ \hline
  \object{HD 147889} & 16:25:24.3 & -24:27:56.6 &    $e$     & V &  3.56\pm0.09  & 177\pm0.7    \\
                     &            &             & 2015-05-19 & V &   3.5\pm0.2   &   \nodata    \\ \hline
  \object{HD 170938} & 18:32:37.8 & -15:42:05.9 &    $f$     & V &   3.7\pm0.2   & 119.0\pm1.6  \\
                     &            &             & 2016-08-06 & V &  3.83\pm0.02  &   \nodata    \\ \hline
  \object{HD 172252} & 18:39:39.9 & -11:52:43.0 &    $f$     & V &   4.6\pm0.2   & 148.0\pm1.2  \\
                     &            &             & 2016-08-05 & V &  4.59\pm0.02  &   \nodata    \\
                     &            &             & 2016-08-06 & V &  4.69\pm0.02  &   \nodata    \\ \hline
  \object{HD 298383} & 09:22:29.8 & -52:28:57.3 &    $b$     & V &  5.23\pm0.01  & 148.6\pm0.05 \\
                     &            &             & 2016-01-30 & V &  5.31\pm0.04  &   \nodata    \\
                     &            &             & 2016-01-31 & V &  5.34\pm0.02  &   \nodata    \\
                     &            &             & 2016-12-22 & V &  5.40\pm0.04  &   \nodata
\enddata
\tablerefs{$a$ \citet{1992AJ....104.1563S}; $b$) \citet{1990AJ.....99.1243T}; $c$) \citet{1975ApJ...196..261S}; $d$) \citet{1982ApJ...262..732H}; $e$) \citet{1982PASP...94..618B}; $f$) \citet{2000AJ....119..923H}.}
\end{deluxetable*}

\begin{figure}
\centering
\includegraphics[width=0.8\linewidth]{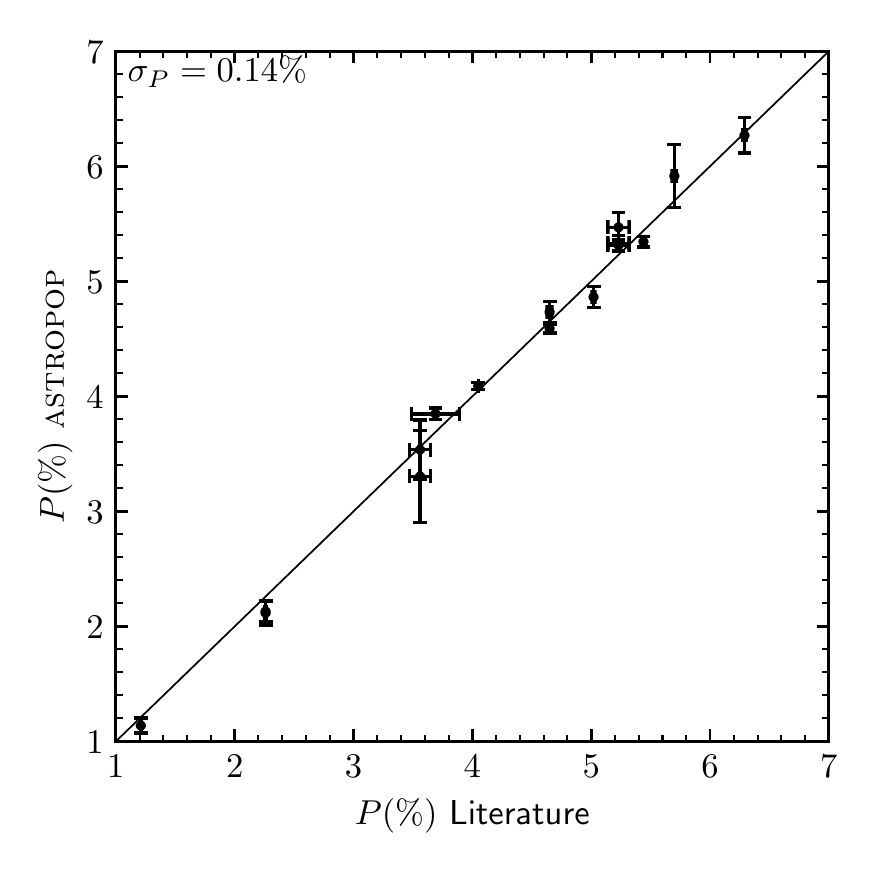}
\caption{Comparison of the polarization level ($P$) obtained from \astropop\ and from the literature (see Table\,\ref{tab:obs_stars}).}
\label{fig:stds}
\end{figure}

Figure\,\ref{fig:stds} compares the polarization level for these stars measured with \astropop\ (SLS method) and the values from the literature, both shown in Tab.~\ref{tab:obs_stars}. These values are in good agreement , giving a dispersion of $\mathrm{rms}=0.16$\%. For most of the analyzed stars, the measurements agree with the literature within 1$\sigma$ errors, and the other agree in 2$\sigma$ errors. Using the equality as a fit, a reduced chi-squared of $\chi_\nu=2.6$ is found, implying systematic differences of similar scale to the statistical error bars.

\subsection{Comparison with \pccdpack}

\pccdpack\ is, at this moment, the main polarimetry reduction package used by IAGPOL users. To verify how the results produced by the two codes behave, synthetic datasets of images -- like those produced by IAGPOL -- by were generated and reduced using both \pccdpack\ and \astropop.\footnote{The code used to generate the synthetic images is on-line available as an Jupyter notebook at \url{https://github.com/juliotux/astropop/blob/master/docs/tutorials/astropop_polarization_models.ipynb}.}

In total, 5 datasets with 400 simulated Moffat-like stars each were generated. The coordinates of the stars were randomly distributed inside the images, with random flux values produced using an exponential distribution, with a flux range from 0 to 500\,000 counts. All the stars had the same value of polarization ($P=10\%$) and the same PA ($\Theta=15\arcdeg$). Each dataset was generated with 16 retarder positions with 22.5\arcdeg step, just like IAGPOL. The image parameters were chosen to mimic the characteristics of the IAGPOL mounted at BC telescope, with a $2048\times2048$ pixels detector and 5~pixels FWHM seeing.

\begin{figure*}
\centering
\includegraphics[width=\linewidth]{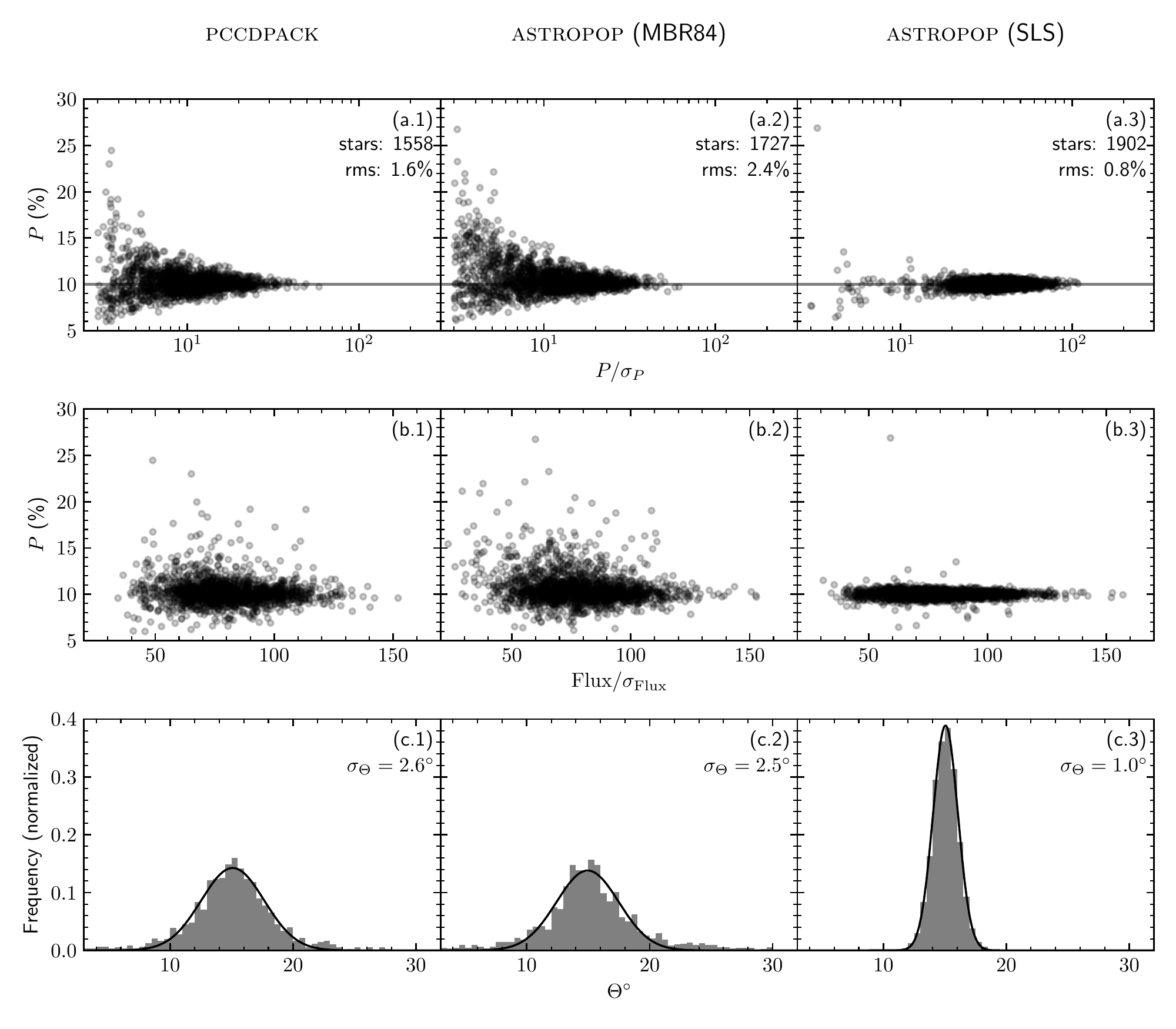}
\caption{Comparison between \pccdpack\ and the two polarimetry calculation modes from \astropop, using simulated data. All the points were simulated at different SNR, but with the same polarization level ($P=10\%$) and PA ($\Theta = 15\arcdeg$). The first column shows the results obtained with \pccdpack, the second and the third the results from \astropop, using the two different polarization algorithms: MBR84 algorithm in second column and by SLS in the third. In the {a.1}, {a.2} and {a.3} panels it is remarked the number of stars with $P/\sigma_P > 3$ and the RMS between the obtained and the simulated values and in the {c.1}, {c.2} and {c.3} panels it is annotated the $\sigma$ of the Gaussian fit showed as the black line. \label{fig:synth_pol}}
\end{figure*}

Both codes showed high detection rates in images, with \pccdpack detecting 1699/2000 pairs and \astropop\ detecting and matching 1923/2000 pairs of stars with no human interaction. The results obtained by both codes are analyzed in Fig.~\ref{fig:synth_pol}. \astropop\ with MBR84 algorithm and \pccdpack\ showed very similar results both in polarization versus SNR (Fig.~\ref{fig:synth_pol}, (a) and (b) panels) and $\Theta$ distributions (Fig.~\ref{fig:synth_pol}, (c) panels), with \astropop\ showing a somewhat larger dispersion than \pccdpack\ ($\sigma_\mathrm{MBR84} = 2.4 \%$ versus $\sigma_\mathrm{\pccdpack} = 1.6 \%$). The probable origin of this difference, since the polarization method is the same, is that \astropop\ calculated more stars with $P/\sigma_P > 3$ than \pccdpack\ (1727 in \astropop\ versus 1558 in \pccdpack), resulted in a more populated region of high dispersion ($3 \leq P/\sigma_P \leq 10$).

However, it is very clear that the best results were obtained with \astropop\ with the SLS algorithm. The SLS algorithm achieved the the smallest dispersion in $P$ ($\sigma_\mathrm{SLS}=0.80\%$), the biggest number of stars with $\mathrm{SNR}>3$ (1902 pairs in total) and the most concise calculations of $\Theta$, with half of the standard deviation from the other methods. Only one star was fit with $P > 15\%$ by SLS, as opposed to 25 by \pccdpack\ and 55 by MBR84 algorithm (see Fig.~\ref{fig:synth_pol}~a.3).

The origin of high dispersion points with $P/\sigma_P < 10$ in \pccdpack\ and MBR84 methods is not directly related to faint stars. When we compare the value of $P$ with the flux SNR ($\mathrm{Flux}/\sigma_\mathrm{Flux}$, Fig.~Fig.~\ref{fig:synth_pol}~b panels), we see that the high dispersion points are spread almost all across the SNR range.

\begin{figure}
\centering
\includegraphics[width=\linewidth]{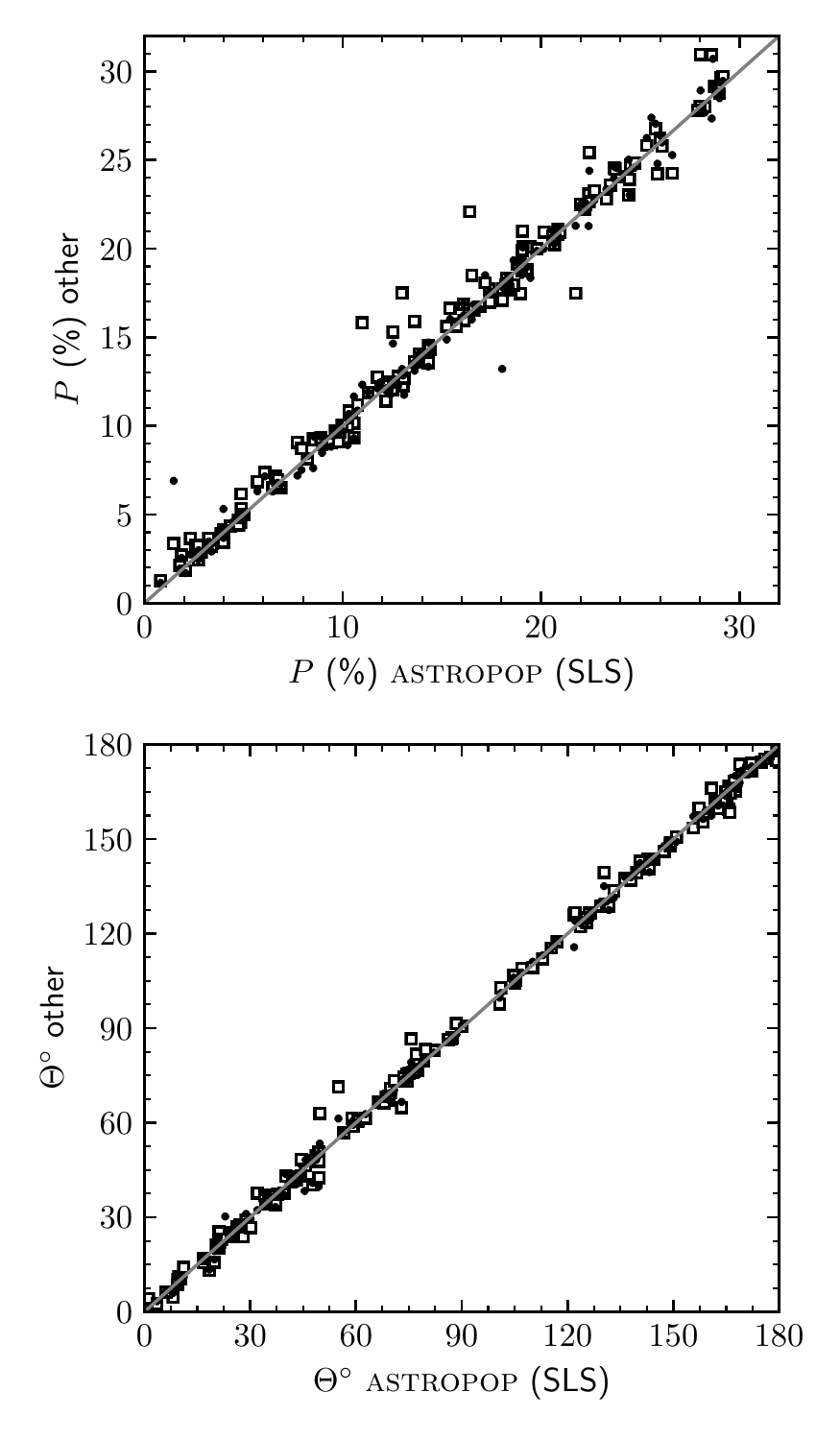}
\caption{Comparison of the values obtained from the \astropop\ SLS method to the \astropop\ MBR84 and \pccdpack\ using simulated data. White squares represent \astropop\ with MBR84 algorithm and black dots represent \pccdpack\ results. Error bars were not plotted due to its randomness according to the star flux. The gray line represents the equality between the values. \label{fig:simutrends}}
\end{figure}

To check the compatibility between the codes across a wide range of $P$ and $\Theta$ values, I generated and reduced a new dataset with random $P$ and $\Theta$ values. Fig.~\ref{fig:simutrends} shows the comparison between \astropop\ results obtained by the SLS algorithm and those from \pccdpack\ and \astropop\ with MBR84 algorithm. No super/underestimation trends were observed at all between the methods. I obtain low reduced chi-squares for all combinations, being around $1.5$ for the $P$ comparison between \pccdpack\ and the two methods of the \astropop\ and around 1.0 when comparing the $P$ obtained by SLS and MBR84 method.

\begin{figure*}
\centering
\includegraphics[width=0.6\linewidth]{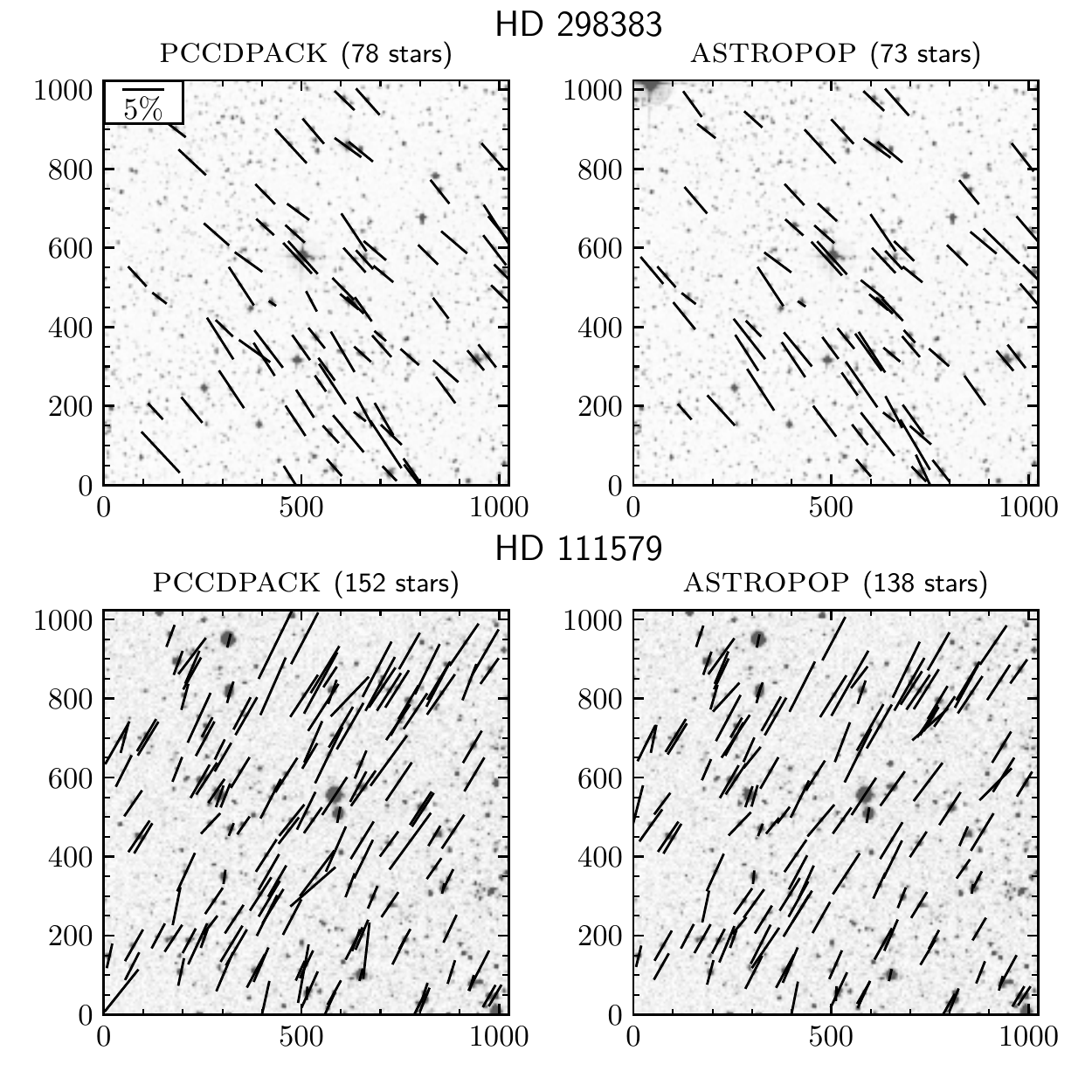}
\caption{Field of HD\,298383 (top) and HD\,111579 (bottom) stars reduced by \pccdpack\ (left) and \astropop\ with SLS method (right). The plotted vectors represent the polarization vectors calculated by both methods and the background images were obtained from DSS2 Red survey from the SkyView service \citep{1998IAUS..179..465M}. Only stars with $P/\sigma_P>5$ are shown and the number of stars present is shown in the panels titles.}
\label{fig:maps}
\end{figure*}

For real data conditions, we analyzed the fields of the standard stars HD\,111579 (observed in $I$ filter in 2010-06-01) and HD\,298383 (observed in $V$ filter in 2016-01-30). Both fields are shown in Fig.~\ref{fig:maps}. \astropop\ reduced each field using SLS algorithm in approximately one minute, including image preprocessing, alignment, polarimetry and photometry reduction. \pccdpack\ took around 2 minutes each field due to the interaction between the user and the program, with no image preprocessing (using the images previously processed by \astropop). \astropop\ detected more stars and grouped more pairs of stars than \pccdpack, identifying 268 pairs in total in the HD\,298383 field and 589 in the HD\,11579 field, versus 197 and 409, respectively, from \pccdpack.

\begin{figure*}
\centering
\includegraphics[width=0.75\linewidth]{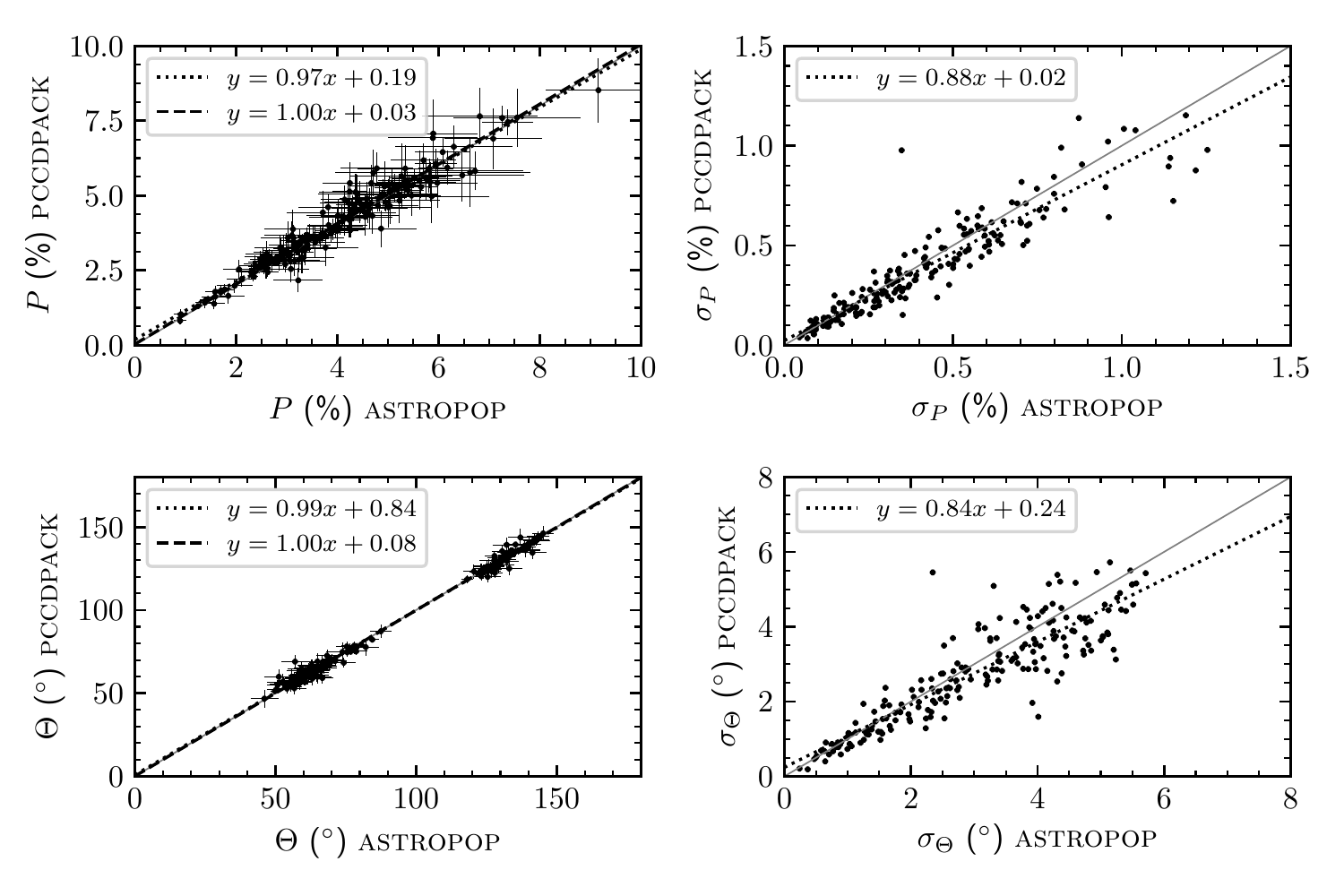}
\caption{Direct comparison between \astropop\ (SLS method) and \pccdpack\ for the HD\,298383 and HD\,111579 fields. In the left panels I show the comparison between the obtained value for $P$ (top) and $\Theta$ (bottom) and the right panels show the errors calculated by the codes. The solid gray lines represent the equality between the values, the dashed lines represent a linear fit with the SNR weighting (only left panels) and the dotted lines the simple linear fit (not weighted).}
\label{fig:realtrends}
\end{figure*}

For the sources measured with both methods, I investigated if there is any systematic trending or error between the methods. Fig.~\ref{fig:realtrends} show the direct comparison between \pccdpack\ and \astropop\ for both $P$ and $\Theta$ values. In total, 192 stars from both fields matched in the two reductions. The presence of the two distinct fields of stars around HD\,298383 and HD\,111579 is clear in the bimodal distribution of $\Theta$ values in Fig.~\ref{fig:realtrends}. The two fields show similar values of $P$.

Linear fits between the points were made using a linear least-squares fit, which fitted parameters (slope and interception) are presented in Table~\ref{tab:fitted}. Two fits were made for each parameter comparison, one weighting the fit by the points SNR, and another not considering the errors in the fit. As we can see, both on the Fig.~\ref{fig:realtrends} and in Table~\ref{tab:fitted}, the fitted slopes are consistent with one and the fitted interceptions are consistent with zero, meaning that there are no systematic error or trend between the parameters obtained by the two reductions. For both $P$ and $\Theta$, the estimated errors associated with \astropop\ are around 10 to 15\% larger than the estimated by \pccdpack. This difference, however, does not affects the number of filtered stars in a relevant way. Lowering the SNR filter of \astropop\ by this value do not changed the number of filtered stars.  The origin of this systematic difference is uncertain. If we compare SLS error calculation method with MBR84, which is theoretically the same as \pccdpack, we find no systematic trend. \astropop\ error estimations are more compatible with the high dispersion measure in Fig.~\ref{fig:synth_pol} and my suggestion is that \pccdpack\ may be underestimating the errors.

\begin{deluxetable*}{cDDDD}
\tablecaption{Parameters of linear fits between \astropop\ and \pccdpack\ reductions shown in Fig.~\ref{fig:realtrends}. \label{tab:fitted}}

\tablehead{\colhead{Parameter} & \twocolhead{Slope} & \twocolhead{Interception} & \twocolhead{Slope} & \twocolhead{Interception}}
\decimals
\startdata 
                & \multicolumn{4}{c}{Weighted Fit} & \multicolumn{4}{c}{Non-weighted Fit} \\
$P$             & 1.00\pm0.01   & 0.03\pm0.04      & 0.97\pm0.02   & 0.19\pm0.08          \\
$\Theta$        & 0.998\pm0.002 & 0.1\pm0.3        & 0.991\pm0.005 & 0.8\pm0.5            \\
$\sigma_P$      & \nodata       & \nodata          & 0.88\pm0.03   & 0.02\pm0.01          \\
$\sigma_\Theta$ & \nodata       & \nodata          & 0.83\pm0.03   & 0.2\pm0.1            \\
\enddata
\end{deluxetable*}

Calculating the reduced chi-squared for $P$ and $\Theta$ correlations, we got $\chi^2_{\nu,P}=0.68$ and $\chi^2_{\nu,\Theta}=0.52$. This means that the dispersion of the points is smaller than statistical errors and the two analyzed codes are in good agreement. Using the two samples Kolmogorov--Smirnov statistical test (KS test) in order to compare the two distributions, I obtain very high $p$-values -- of $p_P=0.90$ and $p_\Theta=0.99$ -- and very low KS-statistics -- with values of 0.06 and 0.04, respectively. So, the KS test indicates that the null hypothesis (i.e. the distributions are the same) cannot be discarded.

\begin{figure*}
\centering
\includegraphics[width=\linewidth]{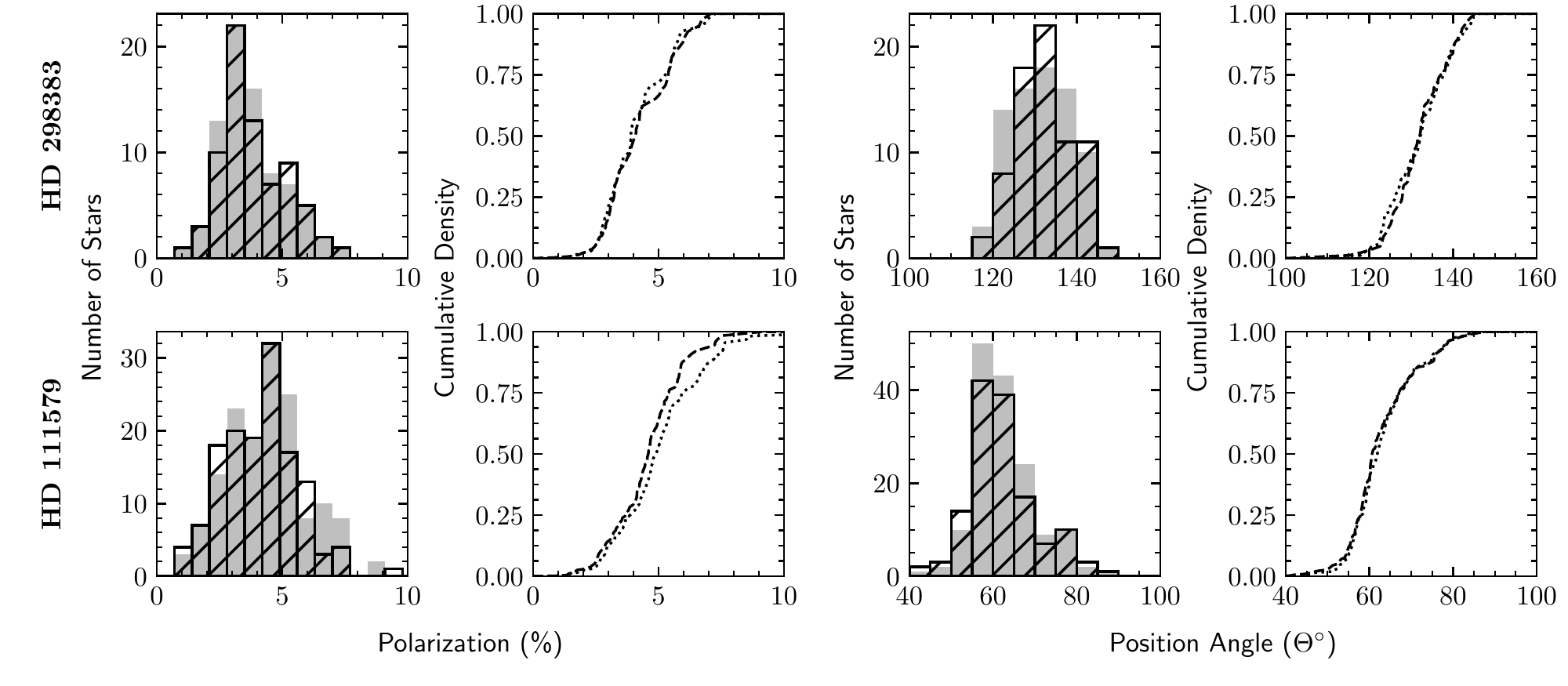}
\caption{Comparative distributions of values obtained from \pccdpack\ and \astropop\ (SLS). All the stars with $P/\sigma_P>5$ are presented. For the histograms, gray bars represent the \pccdpack\ values and hashed bars represent \astropop\ results. For the cumulative distributions, dotted lines represent \pccdpack\ and dashed lines represent \astropop.}
\label{fig:histograms}
\end{figure*}

The distributions of the values for the individual fields are shown in Fig.\ref{fig:histograms}, in which I included all filtered stars from both codes and not just the stars matched in both codes. As we can see, either in histogram analysis and cumulative distribution, even with different number of stars both codes can produce very similar results. The only perceptible difference is the $P$ distribution of the HD\,111579 field, where \pccdpack\ show some more stars with higher level of polarization. The Student's T-test, however, returns a $p$-value of 0.66 for the polarization distribution, implying that the statistical relevance of the difference is small.

With all the different analysis made here, we can conclude that, for polarimetry, \astropop\ can produce results reliable for science, which are compatible with the ones obtained by \pccdpack.

\subsection{Photometric calibration}

\begin{figure}
\centering
\includegraphics[width=\linewidth]{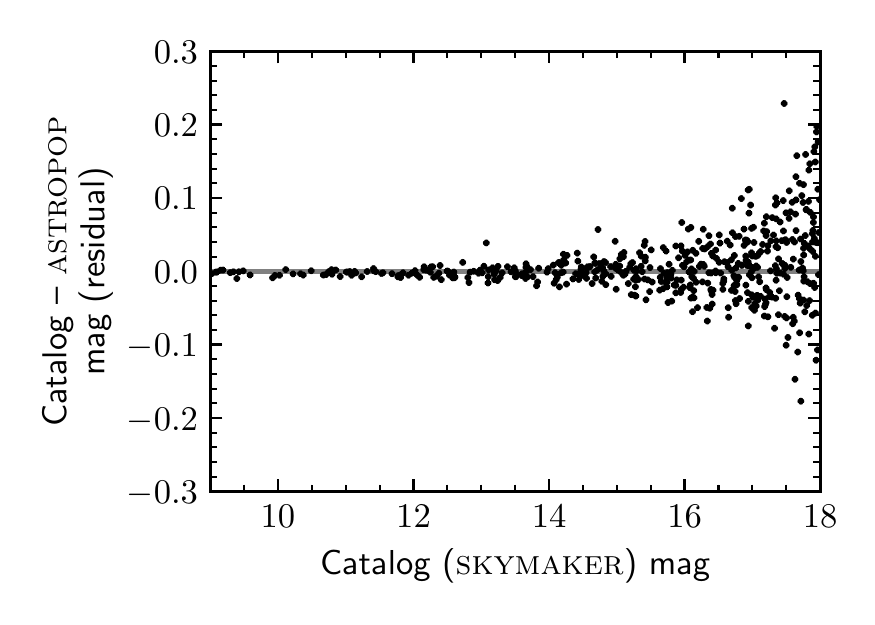}
\caption{Comparison between the reduced and the catalog magnitudes, as a function of the catalog magnitude, of the star-field created with \textsc{skymaker}.
\label{fig:skymaker}}
\end{figure}

The photometric calibration of the code was tested with simulated data, based on a synthetic sky image generated with the Astromatic's \textsc{skymaker} software \citep{2009MmSAI..80..422B, 2010ascl.soft10066B} with the associated catalog of stellar magnitudes. The configurations used in the software was chosen to match the ROBO40 telescope and OPD standard observation conditions (see Appendix~\ref{ap:skymaker}). A detailed comparison between the reduced magnitudes and the magnitudes generated by \textsc{skymaker} is shown in Fig.~\ref{fig:skymaker}.

With this analysis we can see that \astropop\ can reduce the data and determine the magnitudes very precisely having a very low dispersion and residual magnitudes (Fig.~\ref{fig:skymaker}). The rms errors I found are $\sigma=0.002$~mag for stars between 9 and 10~mag, $\sigma=0.006$~mag for stars between 12 and 13~mag and $\sigma=0.08$~mag for stars between 17 and 18~mag. No problems with bad sky subtraction were found in the data. Saturated stars were also discarded automatically by the sharpness in the source detection, with all the dispersion being compatible with theoretical errors.

\begin{figure}
\centering
\includegraphics[width=\linewidth]{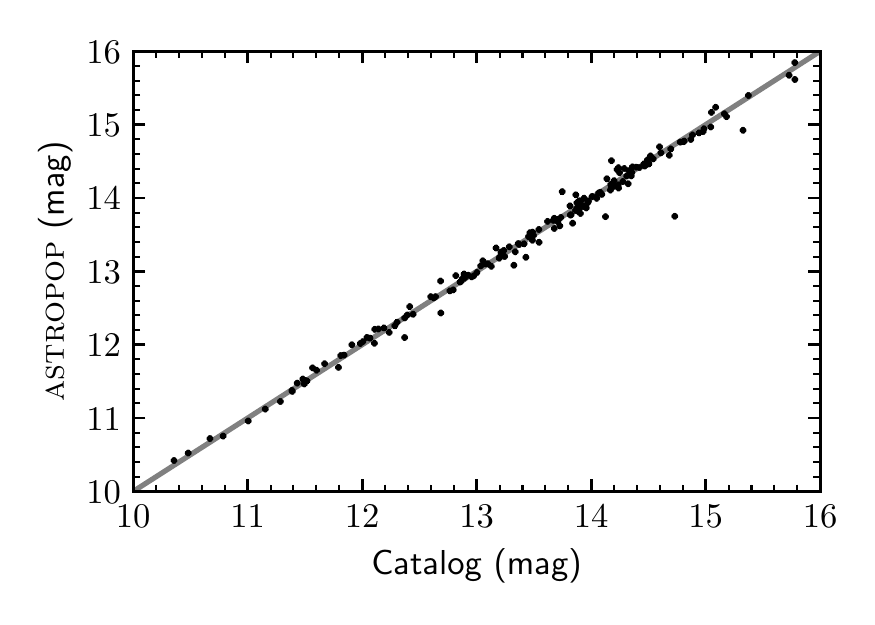}
\caption{Comparison between the catalog and calculated magnitudes with \astropop. For the catalog magnitudes, I used the APASS~DR9 $V$~mag catalog for the field of HD\,298383 (observed in the $V$ filter) field and DENIS $I$~mag for HD\,111579 (observed in $I$). The gray line represents the equality between the two values.\label{fig:realmags}}
\end{figure}

For a real data testing, Fig.~\ref{fig:realmags} shows a comparison between the magnitudes obtained with the \astropop\ reduction for the fields analyzed of HD\,298383 and HD\,111579 and the catalog magnitudes. The field around HD\,298383 was observed in $V$-band, while the field of HD\,111579 was observed using the $I$ band. So, we used two different catalogs for magnitude calibration: APASS~DR9 \citep{2015AAS...22533616H, 2016yCat.2336....0H} for $V$ and DENIS \citep{2000A&AS..144..235C, 2005yCat.2263....0T} for $I$. A small dispersion is observed between the values, with rms=0.13~mag, which is very close to the expected by the catalogs errors and the sources' signal. The systematic errors in this dataset, from the dispersion of the stars in the magnitude range of 10-12, are estimated in 0.05~mag. Only few stars deviated from the expected value which can be variable stars, since no filter of this type of stars was made in the sources.

\astropop\ could produce very reliable values for photometric calibration. As the magnitude calibration depends on source identification in the catalogs, the method of identification of \astropop\ is also tested to produce good results.

\section{Conclusion\label{sec:conclusions}}

In this work, I demonstrated that the \astropop\ reduction package can reproduce polarimetric measurements compared to the literature and to those obtained by the \pccdpack\ code. This means that the code is suitable to perform data reduction with science quality and precision, even for large amounts of data. At this time, the polarimetric results of this code are being used for science production in a paper in preparation \citep{campagnolo.inprep}. In addition, this pipeline is being tested with the ROBO40 telescope and it has the potential to be used in situ to automatically reduce the data produced each night.

The great advantages of \astropop\ over other available codes are:
\begin{itemize}
\item Portability: the code is written in pure-Python, with no need of \iraf\ or any other specific software that has a difficult installation process for the users. It can be installed even under Anaconda environments, with the automatic installation of dependencies;

\item Versatility: \astropop\ is suitable to be used with many instruments from different telescopes without changes in the code, just providing the proper entry parameters;

\item Modularity: the code is modular and any user can easily write its own reduction recipe using the \astropop\ functions;

\item Automatic Reduction: \astropop\ has its own built-in recipes for some cameras (with or without polarization), that can be used for similar instruments without modification of the code.
\end{itemize}

The next steps of \astropop\ development include: (i) the implementation of new methods of astrometric solution, (ii) the solution of circular polarimetry from quarter-wave retarder polarimeters, (iii) the treatment for data from single-beam polarimeters, (iv) the solution for extended objects. In the future, I also intend to include user interfaces (in a initial stage of development) and also quick-look tasks (for quick check the data just after the observation), together with better and more standardized output catalogs.

\acknowledgments
J.C.N.C acknowledges CAPES for the PhD grant. I special thank for Dr. Claudia Vilega Rodrigues for all the theoretical support about polarimetry and data reduction, essential to the development of the code. I thank Dr. Carlos Guerrero Pe\~na for the paper revision and my PhD advisor Dr. Marcelo Borges Fernandes for the useful comments about the paper. I also thank the anonymous referee for the comments that improved the paper during the revision. I also thank the Observat\'orio Pico dos Dias staff for the support during observations nights. This research has made use of the VizieR catalogue access tool (CDS, Strasbourg, France). The original description of the VizieR service was published in A\&AS 143, 23. This work use images from the NASA Skyview service. I also thank for all the open source developers, especially scientific softwares, which work is essential for the science building today. Knowledge has to be free.

\vspace{5mm}
\facility{LNA:BC0.6m}
\software{Astropy \citep[\url{https://astropy.org},][]{2013A&A...558A..33A, 2018arXiv180102634T},
         SkyMaker \citep{2010ascl.soft10066B},
         Astrometry.net \citep[\url{https://astrometry.net},][]{2010AJ....139.1782L},
         Astro-SCRAPPY \citep[\url{https://github.com/astropy/astroscrappy},][]{astroscrappy_repo},
         pccdpack \citep{1996ASPC...97..118M},
         SciPy~stack (\url{https://https://scipy.org/})}

\bibliographystyle{aasjournal}
\bibliography{referencias.bib}

\appendix

\section{\textsc{skymaker} Configure File\label{ap:skymaker}}

The following non-default parameters were used in the \textsc{skymaker} configure file to generate the synthetic image used for photometric testing:

\begin{verbatim}
IMAGE_SIZE      4096            # Width,[height] of the output frame
GAIN            1.9             # gain (e-/ADU)
WELL_CAPACITY   0               # full well capacity in e- (0 = infinite)
SATUR_LEVEL     65535           # saturation level (ADU)
READOUT_NOISE   1.0             # read-out noise (e-)
EXPOSURE_TIME   50.0            # total exposure time (s)
MAG_ZEROPOINT   20.0            # magnitude zero-point ("ADU per second")
PIXEL_SIZE      0.45            # pixel size in arcsec.
PSF_TYPE        INTERNAL        # INTERNAL or FILE
SEEING_TYPE     LONG_EXPOSURE   # (NONE, LONG_EXPOSURE or SHORT_EXPOSURE)
SEEING_FWHM     2.0             # FWHM of seeing in arcsec (incl. motion)
M1_DIAMETER     0.4064          # Diameter of the primary mirror (in meters)
M2_DIAMETER     0.127           # Obstruction diam. from the 2nd mirror in m.
ARM_COUNT       0               # Number of spider arms (0 = none)
WAVELENGTH      0.8             # average wavelength analysed (microns)
BACK_MAG        20.0            # background surface brightness (mag/arcsec2)
STARCOUNT_ZP    1000            # nb of stars /deg2 brighter than MAG_LIMITS
STARCOUNT_SLOPE 0.15            # slope of differential star counts (dexp/mag)
MAG_LIMITS      8.0,18.0        # stellar magnitude range allowed
\end{verbatim}

\end{document}